\documentclass[runningheads]{llncs}
\usepackage{graphicx}
\usepackage{url}

\begin{document}
\title{Find Your ASMR: A Perceptual Retrieval Interface for Autonomous Sensory Meridian Response Videos}

\author{Qi Zhou \and
Jiahao Weng \and
Haoran Xie}
\authorrunning{Q. Zhou et al.}
\titlerunning{Find Your ASMR}
%
\institute{Japan Advanced Institute of Science and Technology, Ishikawa 9231292, JAPAN \\
\email{xie@jaist.ac.jp}}

\maketitle             
\begin{abstract}
Autonomous sensory meridian response (ASMR) is a type of video contents designed to help people relax and feel comfortable. Users usually retrieve ASMR contents from various video websites using only keywords. However, it is challenging to examine satisfactory contents to reflect users' needs for ASMR videos using keywords or content-based retrieval. To solve this issue, we propose a perceptual video retrieval system for ASMR videos and provide a novel retrieval user interface that allows users to retrieve content according to watching purpose and anticipated expectations, such as excitement, calmness, stress and sadness. An ASMR video perception dataset is constructed with annotations on affective responses after watching the videos. To verify the proposed video retrieval system, a user study is conducted showing that users can retrieve satisfactory ASMR contents easily and efficiently compared to conventional keywords-based retrieval systems.\footnote{This is a pre-print author's version of this article. The final authenticated version is available online at proceedings of HCII 2022(International Conference on Human-Computer Interaction)}

\end{abstract}
\section{Introduction}
Autonomous  sensory meridian response (ASMR) is a type of video content that has become popular in recent years due to the pervasive influence of social medias, such as YouTube and TikTok. Especially for young people, ASMR has changed their life styles with the daily use of applications for relaxing and sleeping. As reported in the previous studies, ASMR videos can bring users the sensation phenomenon called tingles \cite{ASMR2,poerio2018more}. This sensation is mainly felt at the back of the users’ heads and is accompanied by a sense of pleasure and relaxation. However, examining satisfactory ASMR videos from conventional retrieval interfaces remains a challenging issue.   

\begin{figure}[t]
  \centering
  \includegraphics[width=\linewidth]{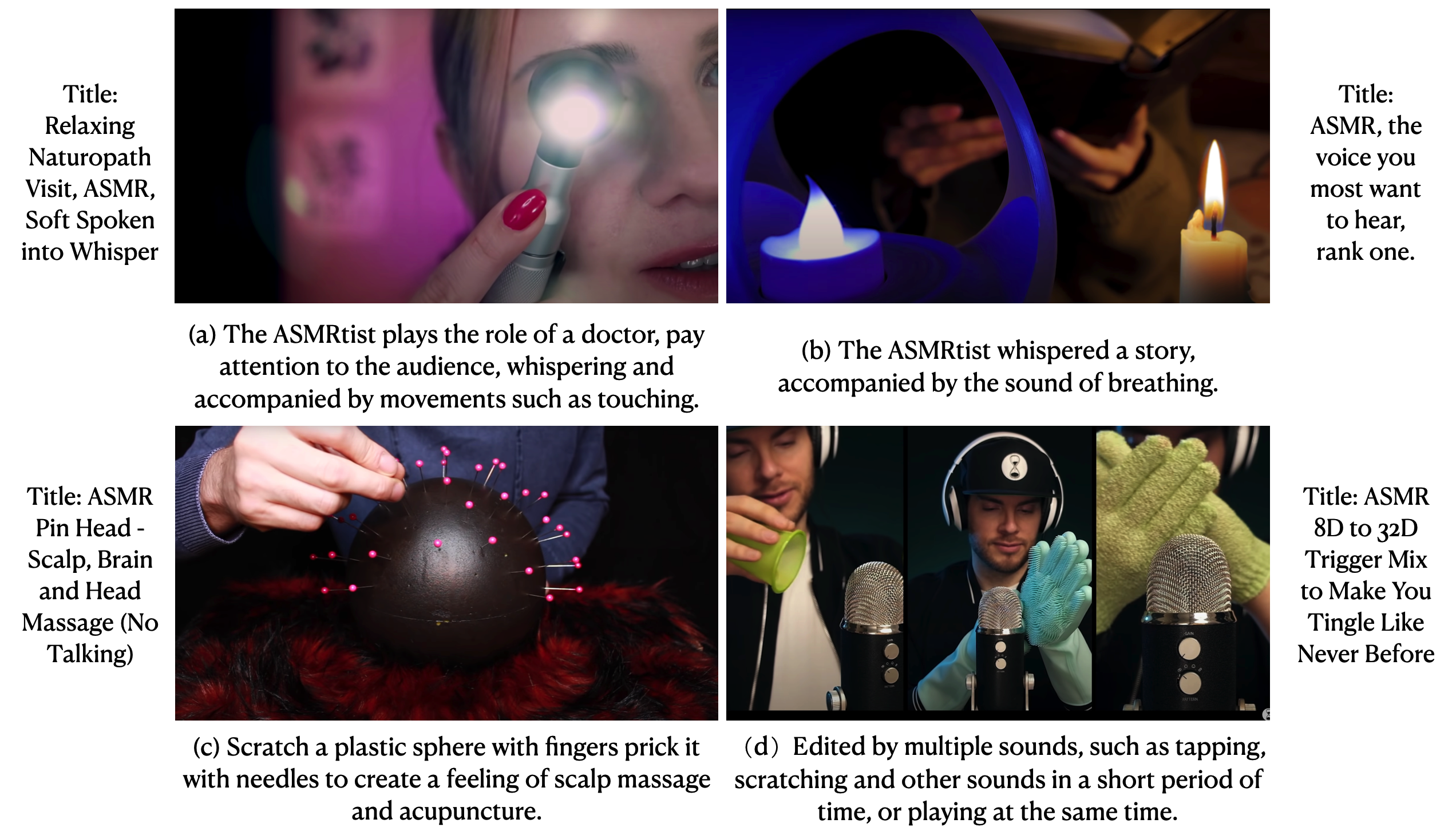}
  \caption{Finding satisfactory ASMR videos using only the keywords in the video titles is difficult. (a)–(d) indicate the four different categories of ASMR videos used in this work.}
  \label{fig:P1_image}
\end{figure}

As shown in Figure~\ref{fig:P1_image}, the titles of ASMR videos usually describe a simple and personal introduction to the video content and have difficulty describing videos for special retrieval purposes, such as relaxation and looking for companionship and attention. The video frames of ASMR videos are usually produced with the action sounds and spoken voices. Classifying videos with multi-modal information is a complicated issue. Therefore,  conventional keywords-based and content-based video retrieval approaches may fail for ASMR videos. In such cases, the users may want to relax but continuously switch ASMR videos because they are dissatisfied with the retrieved results. They wasted time and become unable to relax. 

To solve these issues, we propose a novel perceptual video retrieval interface for ASMR videos. In the proposed system, an ASMR video dataset from online sources was constructed. Then, the participants were asked to evaluate the perception scores of all the collected videos. We implemented a perceptual video retrieval interface with perception filters. We conducted a comparison study between the proposed system and conventional retrieval interfaces to verify the user experience in the user study. 

\section{Related Works}
ASMR is a sensory phenomenon that includes different triggers in videos, such as whispering, personal attention, crisp sounds and slow movements. This sensory– emotional phenomenon can be felt in listeners’ head, neck and shoulders~\cite{ASMR2}. The sensitivity of the triggers can be measured with a resting-state functional magnetic resonance imaging scan~\cite{SMITH2020103021}. ASMR videos have been investigated to study the emotional and physiological correlation with responses~\cite{poerio2018more}. Aside from the tingling sensation, ASMR videos have been reported to relate to experiences of social connection and physical intimacy~\cite{asmr21}. An online community of video sharing was found to have various ASMR videos created by ASMR artists looking for cultural and scientific legitimacy~\cite{SMITH2019}. In this work, we  focus on the video retrieval of ASMR videos.

The common approaches for video retrieval are keyword and content-based video indexing and retrieval~\cite{video20}. A hierarchical retrieval interface was proposed to explore video content through poster-style summarization~\cite{weng22}. A perception-based approach was explored to achieve the desired design using sensation words~\cite{fujita19}. A perceptual video summarization and retrieval was proposed to provide a precise representation of video contents~\cite{peceptual19}. The interaction modalities and parasocial attractions of ASMR videos was annotated for multi-modal video interactions~\cite{niu22}. However, an effective way to explore ASMR videos has been lacking in previous works. To solve this issue, we aimed to achieve perceptual retrieval by annotating the videos in the dataset construction.

\section{Retrieval System}

\subsection{System Overview}
\begin{figure}[t]
  \centering
  \includegraphics[width=\linewidth]{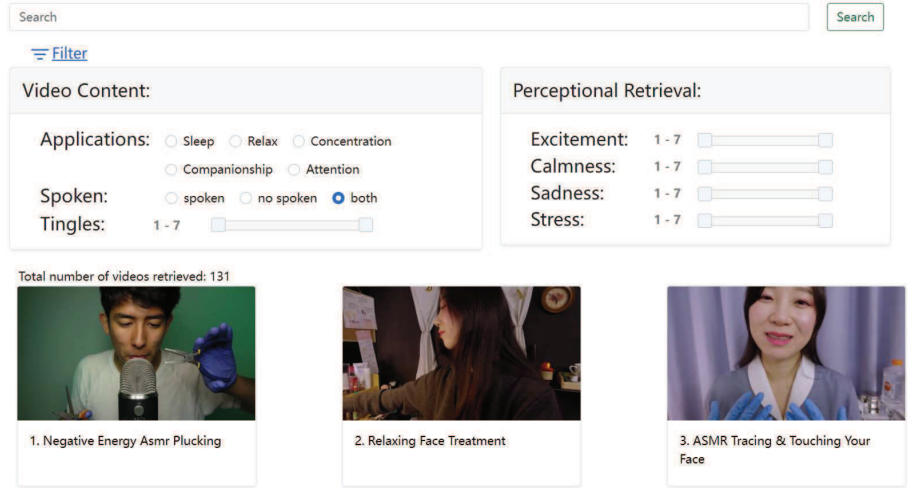}
  \caption{The proposed perceptual video retrieval interface.}
  \label{fig:iu_image}
\end{figure}

The proposed retrieval system provides a perceptual retrieval interface in which users can adjust their perceptual parameters to find satisfactory ASMR videos as shown in Figure~\ref{fig:iu_image}. Users can adjust the application purpose of watching ASMR videos, the desired level of tingles, and the perceptual expectations. For the retrieval system, an ASMR video dataset was constructed, and the participants were asked to evaluate the videos for the level of tingles and perceptual expectations. The proposed system was confirmed to be superior in terms of retrieval time cost and the quality of retrieved videos with satisfactory usability scales.

\subsection{ASMR Video Dataset}
We classified all the collected videos into four video categories according to previous findings~\cite{poerio2018more}: (a) spoken with high interactivity, (b) spoken with low interactivity, (c) no spoken and one or a few contents, and (d) no spoken and  multiple contents. ~\ref{fig:P1_image} shows example videos of the different categories. In total, 131 ASMR videos were collected from YouTube: 41 videos for (a), 29 videos for (b), 36 videos for (c), and 25 videos for (d). The representative content of each video was selected manually and cut into 3–5 min video clips.

\subsection{Perceptual Annotation}
For all the collected videos in the dataset, the ASMR videos were annotated with the following perception metrics: (a) Anticipated applications: Five purposes of watching ASMR videos from the typical scenarios of watching ASMR videos were presented: sleep, relaxation, concentration, companionship, and attention. (b) Tingles: This refers to the degree of stimulating effect of the ASMR video. (c) Perceptual expectation: This refers to positive effects on human emotions. 

According to a previous study \cite{poerio2018more}, the perceptual expectations were divided into four categories, namely excitement, calmness, sadness, and stress, which were used to measure users' perceptual expectations after watching the ASMR video. We adopted these four expressions for the video annotation.

We asked four participants (female graduate students) to join our video annotation. The participants were asked to watch the ASMR videos and annotated them using the aforementioned perception metrics. For each video, the participants were asked to answer the following six questions on a seven-point Likert scale, with 1 indicating strong disagreement and 7 indicating strong agreement.
\begin{itemize}
\item[Q1] Can you feel tingles from this ASMR video?
\item[Q2] Do you feel more excited after watching the video?
\item[Q3] Do you feel calmer after watching the video?
\item[Q4] Do you feel sadder after watching the video?
\item[Q5] Do you feel more stressed after watching the video?
\item[Q6] For which purpose do you think this ASMR video is suitable? (multiple-choice questions)
\end{itemize}

According to a survey of the participants (works) in advance, all four participants had experience watching ASMR videos. They had different preferences for ASMR, with two of them liking humans speaking in ASMR videos and the others liking ASMR videos without speaking. Moreover, two participants were more satisfied with videos that made people calmer, and the other two were more satisfied with videos that made people more excited. All videos were randomly assigned to the participants according to the four categories (Figure \ref{fig:P1_image}).

\subsection{User Interface}
As shown in Figure~\ref{fig:iu_image}, we designed a novel perceptual video retrieval interface with a perception filter for an advance search. For the filer, we created two sections, namely, the video content section and the perceptional retrieval section, to fulfill the user search scenario. The video content section has three items: applications, spoken, and tingles. Application denotes retrieval for the specific purpose of watching, and spoken denotes searching for videos that have or do not have vocals. Users employ a two-handle slider to select the range of the tingle feature. The perceptional retrieval section has four items: excitement, calmness, stress, and sadness. These items also use a two-handle slider for range selection.

\begin{figure}[htb]
  \centering
  \includegraphics[width=0.9\linewidth]{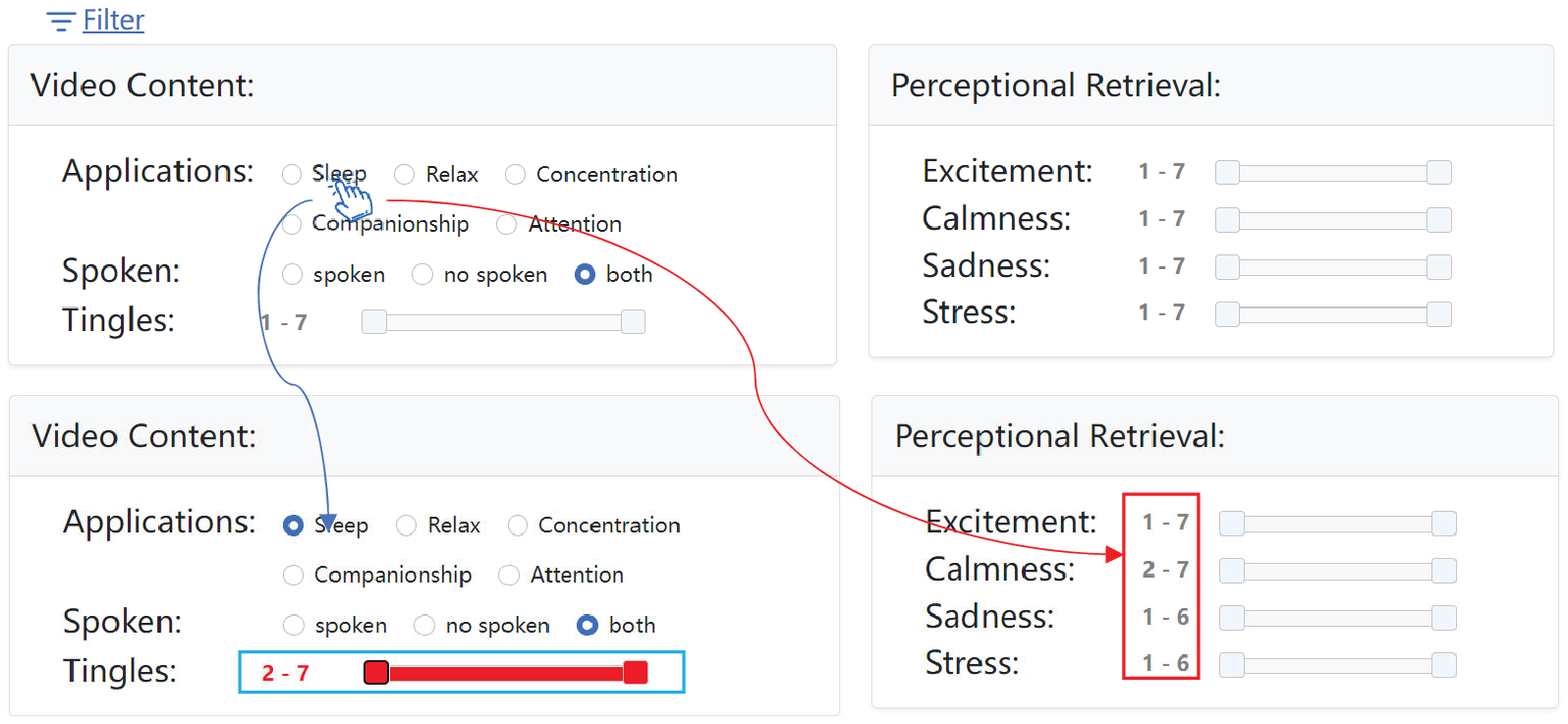}
  \caption{Workflow of the perceptual filter in the proposed user interface.}
  \label{fig:cui_image}
\end{figure}

When users chooses one watching application, the system automatically determines the maximum and minimum video perception values in the dataset and changes the corresponding values on the sliders simultaneously, as shown in the red box in Figure~\ref{fig:cui_image}. The default states of the sliders are illustrated in Figure~\ref{fig:iu_image}, When users change the range of a specific item, the slider can be changed to the activated state, as shown in the blue box in Figure~\ref{fig:cui_image}.

\begin{figure}[]
  \centering
  \includegraphics[width=0.9\linewidth]{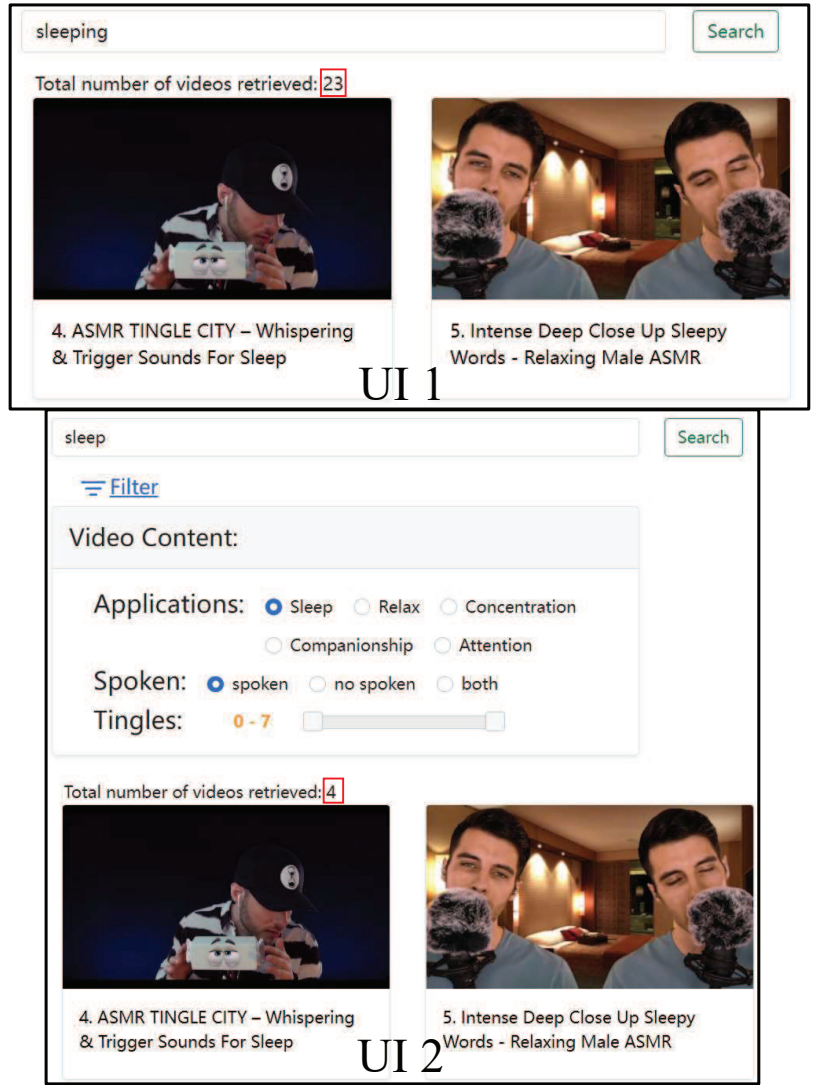}
  \caption{Conventional video retrieval interfaces used in the comparison study. UI-1 represents keyword-based retrieval, and UI-2 represents content-based retrieval.}
  \label{fig:uui_image}
\end{figure}
\section{User Study}

\subsection{Comparison Study}

We conducted a comparative study between the proposed perceptual retrieval interface and two conventional retrieval approaches: keyword-based video retrieval (UI-1) and content-based video retrieval (UI-2). These two user interfaces were implemented for the user study, as shown in Figure~\ref{fig:uui_image}. UI-1 denotes a traditional user interface that only has a keyword search, and UI-2 has has a combined traditional user interface and the video content section of the proposed system. 

All participants were asked to use each of the three video retrieval interfaces to complete three different assigned tasks. Our task was designed based on a specific scene, the content of the video, and the perceptual expectations after watching the video. For example, “You are going to rest and sleep. Please find as many suitable ASMR videos as possible (no spoken, used for relaxation, and can make people calmer or more excited).”

We invited six graduate students (three males and three females), all of whom familiar with ASMR videos, to join the study. The participants were randomly assigned retrieval and permutation tasks. After the participants retrieved the ASMR videos, they were asked to watch them in order, from beginning to end, and confirm whether they were satisfactory. The amount of time the participant took to find the first satisfactory video and the time interval between finding two satisfactory videos were set as the criteria for evaluating the proposed system. The total number of video views and the number of satisfaction videos for each participant were also recorded.

\subsection{User Experience Study}

We asked six graduate students (two males and four females), all of whom were familiar with ASMR videos, to join the user experience study. They were asked to use the proposed retrieval system to complete two tasks:
\begin{itemize}
\item[(1)] Find as many ASMR videos as possible, with constraints of no spoken, and for relaxation and calming applications.
\item[(2)] Find satisfactory ASMR videos easily.
\end{itemize}

After completing each task, the participants watched the retrieved videos for 5-10 min to determine whether they were satisfied with the retrieved results. They were then asked to answer question items from the system usability scale (SUS).

\section{Results}

\begin{figure}[t]
  \centering
  \includegraphics[width=0.7\linewidth]{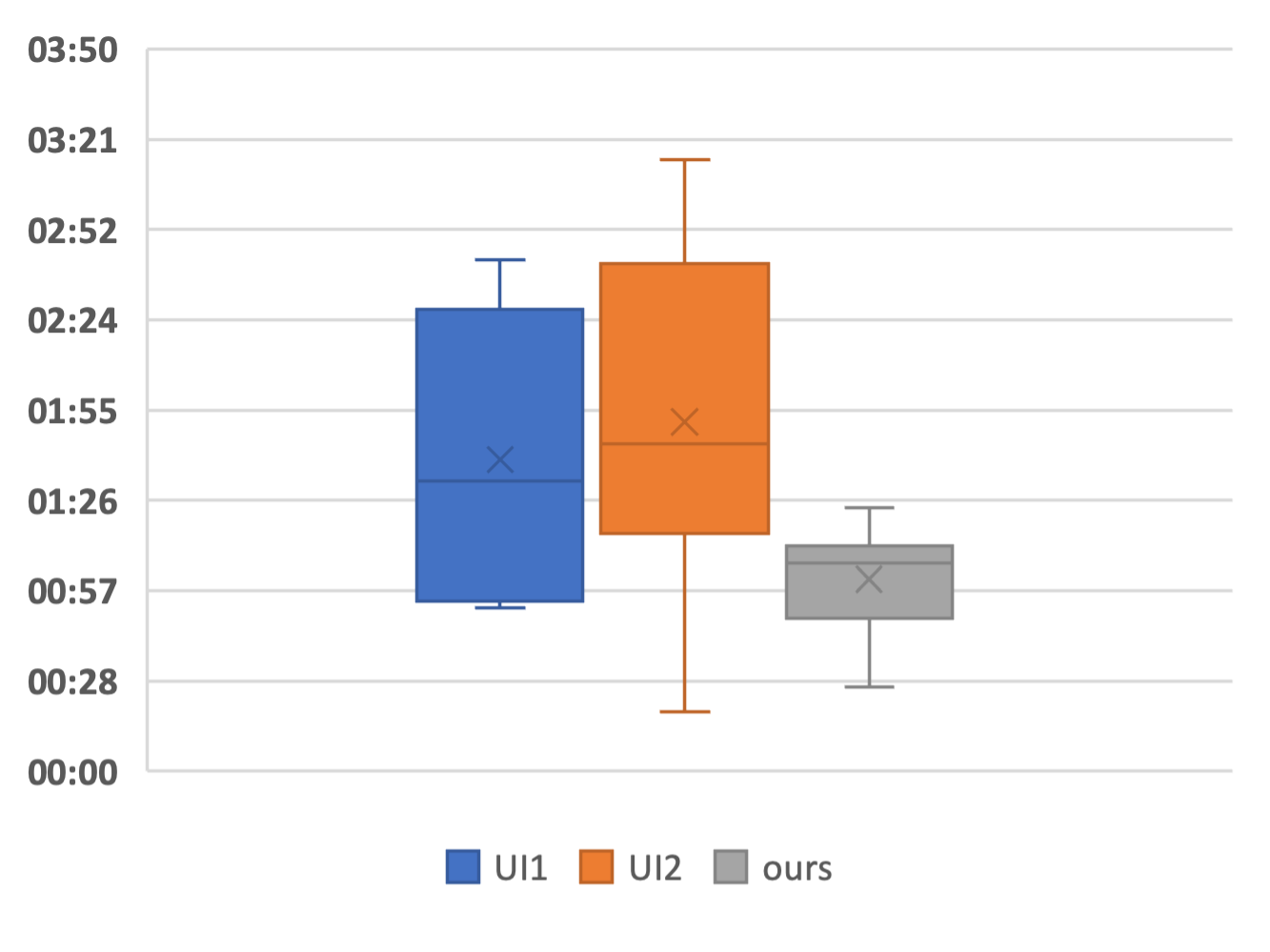}
  \caption{Amount of time for finding the first satisfactory video.}
  \label{fig:C1_image}
\end{figure}

\subsection{Implementation Details}
This system was implemented on Python, and used a Django 3.2.0 web framework. Bootstrap 4.6.1, an open source cascading style sheets (css) framework, was used to render and decorate the website. JavaScript Query 3.6.0 (jQuery), an open-source JavaScript library, was used to implement user interactions. When users chose one watching purpose of the application, we used the asynchronous JavaScript and XML function in jQuery to obtain specific values from the dataset and modify the values in the sliders without refreshing the website. 

A two-handle slider (Figure~\ref{fig:cui_image}) was designed with a combination of two traditional (only one handle) sliders. The z-index, a css attribute, was used to overlap the two traditional sliders. Then, we declared a function to determine if the two sliders would collide. If the left handle had the same value as the right one, it would collide. In this case, the left handle would not be able to move to the right anymore, and vice versa.

\subsection{Comparison Study}

Figure~\ref{fig:C1_image} shows that the proposed interface could find satisfactory videos for all participants in the shortest amount of time. Although our interface took more time in the initial parameter setting than a conventional keyword retrieval, more accurate retrieval results were achieved in a significantly shorter amount of time for the participants to find the first satisfactory video. When using the keyword retrieval method, some participants found the first satisfactory video faster because they could quickly locate familiar video content by inputting keywords. The content-based retrieval method (UI2) disregarded the emotional needs of the participants. Thus, compared to the proposed perceptual retrieval system, the variance in the video content distribution was larger, and the participants have a relatively lower probability of quickly finding the first satisfactory video.

\begin{figure}[htb]
  \centering
  \includegraphics[width=0.7\linewidth]{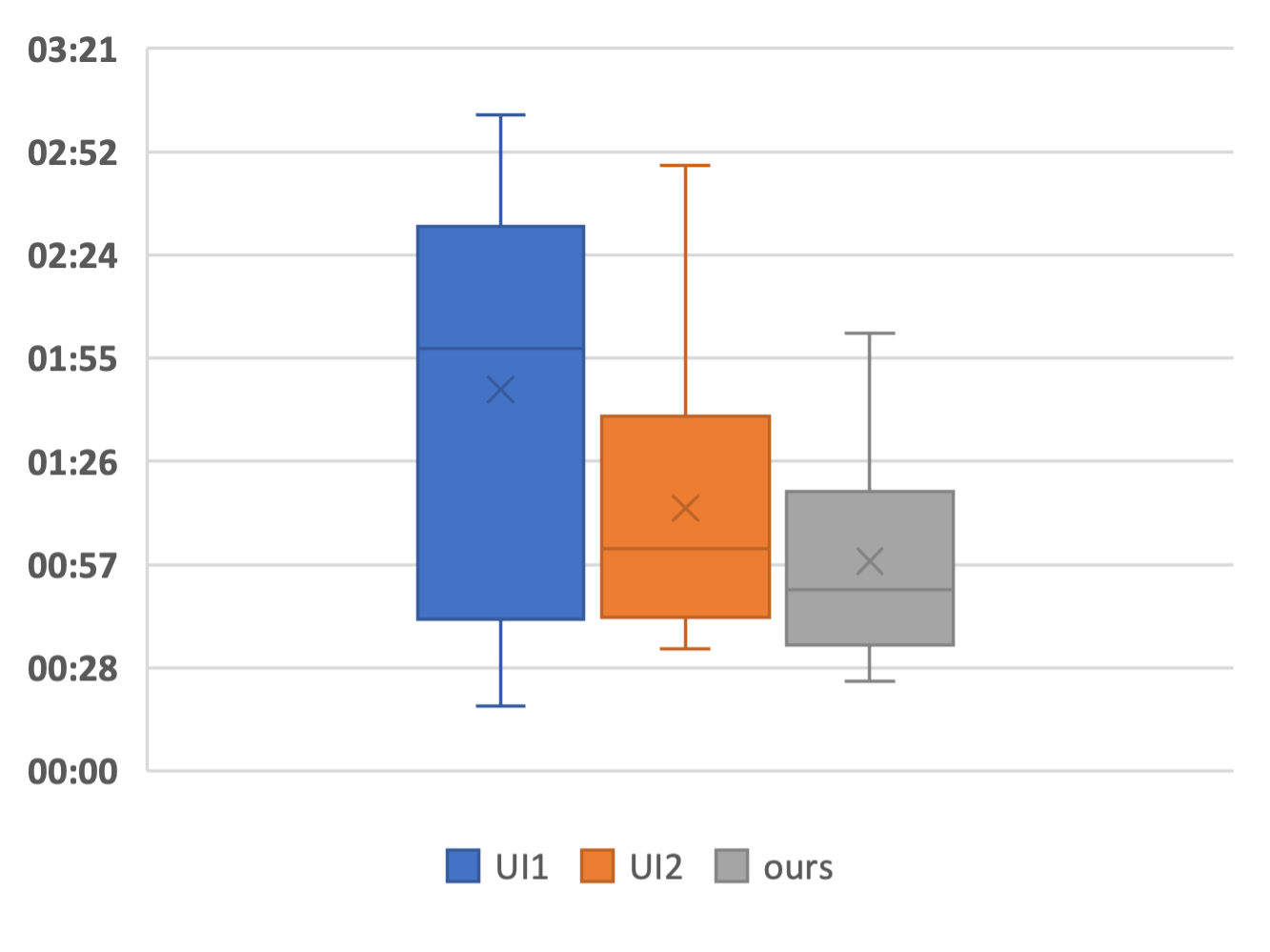}
  \caption{Time intervals between finding two satisfactory videos.}
  \label{fig:C2_image}
\end{figure}

As shown in Figure~\ref{fig:C2_image}, a short interval was observed between finding two satisfying videos consecutively when using the proposed interface. The proposed system was verified to retrieve more satisfactory videos, and users had a better experience in finding and switching between satisfactory ASMR videos. The participants revealed that they usually did not watch ASMR videos completely. When bored with the content of an ASMR video, they would look for other videos and switch between them until they find satisfactory video content. Therefore, we consider that the interval between closing one video and finding another satisfactory one is an important indicator when evaluating a retrieval system, as a short interval can reduce the energy loss caused by switching between videos and significantly increase the comfort of system users. One of the participants commented the following about the proposed retrieval system was: “I feel that the retrieved videos were very new and that most of them were very interesting. I can quickly find the next satisfactory video.”

\begin{figure}[htb]
  \centering
  \includegraphics[width=0.7\linewidth]{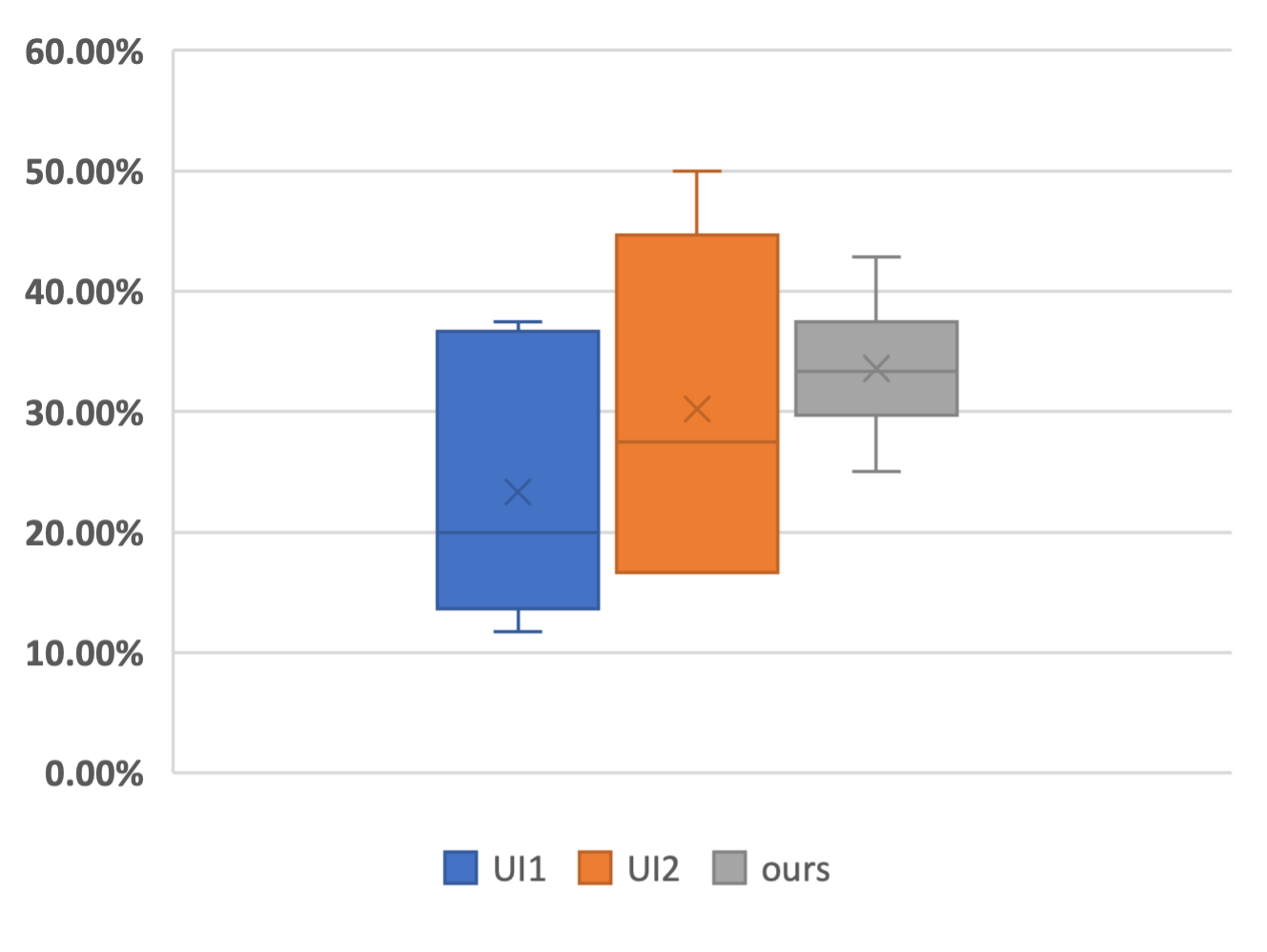}
  \caption{Ratio of satisfactory videos to total viewed videos.}
  \label{fig:C3_image}
\end{figure}

As shown in Figure~\ref{fig:C3_image}, the proposed system can retrieve the highest ratio of satisfactory videos relative to the total number of viewed videos.  The participants indicated that the proposed retrieval method increased the diversity of video types retrieved. According to one of the participants, "I usually use a few keywords I know to search for ASMR videos, In this way, I can find videos that meet my preferences. However, after a period, the videos retrieved using these keywords will have no new content, and most of them I have already seen. However, your retrieval method is not limited to specific video content, and it can find many videos with unexpected new content."

\subsection{User Experience}
\begin{figure}[htb]
  \centering
  \includegraphics[width=0.9\linewidth]{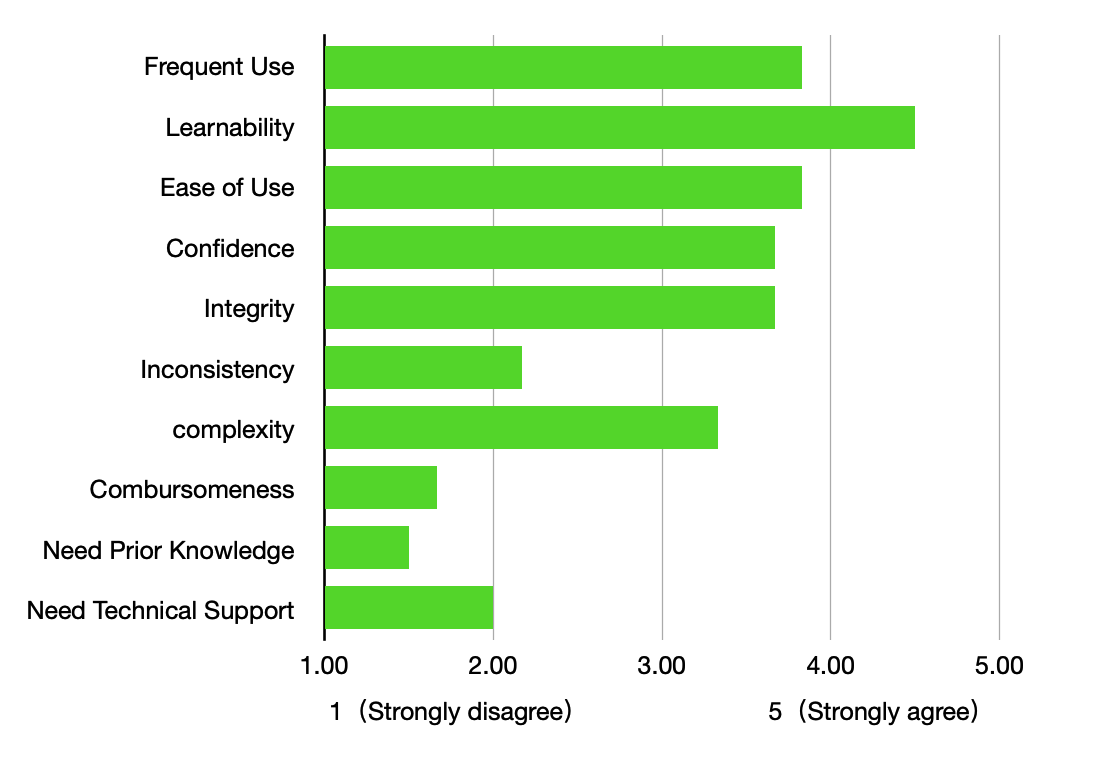}
  \caption{Mean SUS responses from the user study.}
  \label{fig:UE1_image}
\end{figure}

The results of System Usability Scale (SUS) questions are shown in Figure~\ref{fig:UE1_image}. Five participants reported that they were willing to use our retrieval system regularly confident in using it. They found the proposed system easy to learn. However, four participants reported that the system design was complex. We intend to improve our interface design in our future work. The average score of the proposed retrieval interface was 72.08 (out of 100), which implies good overall usability.

\section{Conclusion}
We proposed an ASMR video retrieval interface based on user perceptions. We collected the annotation data of an ASMR video dataset for watching purposes and perception of feelings. In contrast to the conventional keyword- and content-based retrieval interfaces, the proposed retrieval interface achieved better retrieval quantity and accuracyat less time. Through the system usability experiment, the proposed system was verified to achieve good overall usability. 

In future work, we intend to increase the number of participants in the user study and the number of participants and videos during video annotation. Using a deep supervised learning approach for the estimation of perceptual metrics of an unknown ASMR video using the annotation dataset is a promising topic. The proposed interface design can be improved for simplicity of use. We also plan to examine the use of a perception-driven interface for other daily activities, such as taking selfies~\cite{Fang2018SelfieGS} and animation design~\cite{motion21}.

\section*{Acknowledgement}
The authors thank the anonymous reviewers for their valuable comments. This project has been partially funded by JAIST Research Grant and JSPS KAKENHI grant JP20K19845, Japan.

\bibliographystyle{splncs04}
\bibliography{reference}
\end{document}